\documentclass[sigconf]{acmart} 

\AtBeginDocument{%
  }

\renewcommand\footnotetextcopyrightpermission[1]{}
\usepackage{enumitem}
\usepackage{multirow}
\usepackage{graphicx}
\usepackage{adjustbox}
\usepackage{subcaption}

\begin{document}

\title{Schemex: Discovering Design Patterns from Examples through Iterative Abstraction and Refinement}

\author{Sitong Wang}
\affiliation{
  \institution{Columbia University}
  \city{New York}
  \state{NY}
  \country{USA}
}
\email{sw3504@columbia.edu}

\author{Lydia B. Chilton}
\affiliation{
  \institution{Columbia University}
  \city{New York}
  \state{NY}
  \country{USA}
}
\email{chilton@cs.columbia.edu}

\renewcommand{\shortauthors}{Wang et al.}


\begin{teaserfigure}
\centering
\includegraphics[width=0.89\textwidth]
{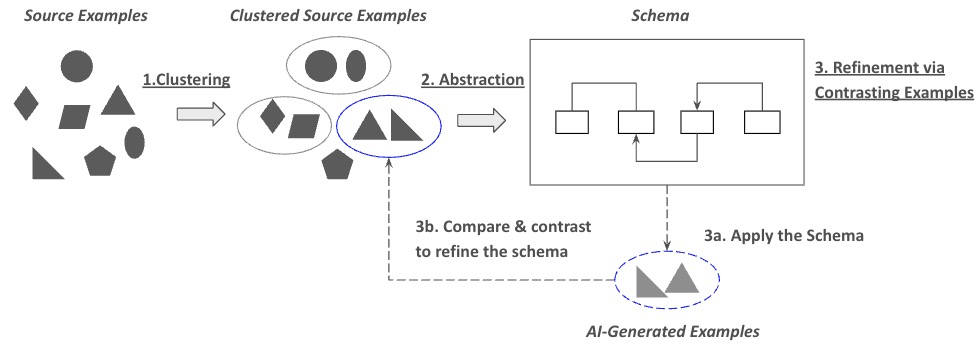}
    \caption{Schemex is an AI-powered workflow that helps users induce actionable schemas from examples. It comprises three key stages: clustering, abstraction, and refinement via contrasting examples.}
    \label{fig:teaser}
\end{teaserfigure}

\maketitle
\pagestyle{plain} 

\section{Introduction}

Expertise is often built by learning from examples.
Chefs study recipes to understand flavor principles, engineers reverse-engineer products to uncover key design concepts, and musicians analyze compositions to master harmonic structures.
This process—schema induction~\cite{gick1983schema}—helps us identify patterns from examples. 
It transforms fragmented observations into usable knowledge: a chef creates new dishes, an engineer solves unexpected problems, and a musician reinvents genres.
As the number and complexity of examples increase, schema induction becomes essential for synthesizing overwhelming information and extracting meaningful insights.

Despite its importance, schema induction remains a challenging cognitive task.
While well-established domains like music composition benefit from expert-created schemas (e.g., chord progressions), emerging fields—such as creating news TikToks~\cite{reelframer}—lack such structured frameworks.
Due to the tacit nature of knowledge, even creators of examples often struggle to articulate the underlying schemas.
The process involves detecting latent patterns across numerous examples, abstracting generalizable rules, and iteratively refining schemas as new information emerges.
Yet, current sensemaking tools focus more on organizing examples than on actively supporting the insight-generation process.

Recent advances in generative AI reasoning capabilities, particularly DeepSeek R1's ``slow thinking''~\cite{guo2025deepseek}, offer new opportunities for supporting schema induction through human-AI collaboration.
We present Schemex, an AI-powered workflow that augments human schema induction through three stages:
\begin{enumerate}[noitemsep]
\item Clustering: Grouping examples by latent similarities
\item Abstraction: Extracting structural patterns within clusters
\item Refinement via contrasting examples: Sharpening schemas through AI-generated contrasting examples
\end{enumerate}
The process forms a collaborative loop in which AI performs clustering, abstraction, and contrastive learning that significantly reduce the cognitive load of humans.
Meanwhile, humans evaluate and refine the outputs, controlling the iteration cycles.

We conducted an initial evaluation of Schemex through two real-world case studies:
\begin{enumerate}[noitemsep]
\item Analysis of HCI Paper Abstracts: Assisting researchers in analyzing abstracts from CHI best papers
\item Multimodal Analysis of News TikToks: Helping journalists understand the schemas of trending news TikToks
\end{enumerate}
Qualitative analysis demonstrates the high accuracy and usefulness of the generated schemas.

Overall, we explore the potential of AI as a cognitive collaborator in schema induction. 
Future work will focus on exploring more flexible ways for workflow construction to incorporate key components such as abstraction and iteration. 
This involves automating low-level operations, such as determining the necessary number of iterations, to allow humans to concentrate on high-level thinking.

\section{Schemex Workflow}

The Schemex workflow (see Figure \ref{fig:teaser}) assists users in transforming a set of examples into actionable schemas through three AI-assisted stages.
For instance, to obtain schemas for writing HCI paper abstracts, the input can be 20 CHI’24 best paper abstracts.  

Initially, the workflow clusters the 20 abstracts into different groups, where each group has distinguishable features—for instance, separating empirical studies from system papers.  
Next, for each cluster, the workflow examines the examples and derives schemas specific to each cluster, identifying initial schemas like ``\textit{Motivation}-\textit{Problem}-\textit{Method}-\textit{Findings}-\textit{Implications}" for empirical studies and ``\textit{Motivation}-\textit{Problem}-\textit{System}-\textit{Evaluation}-\textit{Design Recommendations}'' for system papers.  
Finally, we refine the schema through contrastive learning: we take the initial schema and ask AI to write abstracts based on it, then ask AI to compare the generated abstracts with the original human-authored ones and iterate the schema.  

The resulting schemas are actionable and comprehensive, covering a diverse set of examples and providing guidance on how to write abstracts for different genres of HCI papers.
Users can utilize the resulting schemas as a guide to write their own abstracts.

\subsection{Stage 1: Clustering}
Clustering identifies subclasses of examples and is crucial in preventing schema overgeneralization.  
When given a set of examples, it is often unclear if they share the same schema.  
Generalizing across structurally different examples (e.g., empirical study and system papers), can lead to weak and bland schemas.  
Therefore, we must perform clustering first to identify examples that are similar to each other and then find the schema specific to each cluster.

We instruct DeepSeek-R1 to cluster the examples and provide reasoning.
We choose DeepSeek-R1 for its advanced reasoning capabilities.
For our CHI abstract dataset, the model processes all 20 examples simultaneously, producing clusters and explaining the common features that examples within each cluster share.
Users can then easily verify the clustering results.
They can refine the clustering if they find any errors and can also merge clusters or request further division within a cluster if needed.

For instance, DeepSeek-R1 maps the 20 abstracts into 4 clusters:
\begin{itemize}[noitemsep]
    \item Empirical Studies (6 examples)
    \item Theoretical Contributions (4 examples)
    \item System Design and Evaluation (6 examples)
    \item Ethnographic Insights (4 examples)
\end{itemize}
It also provides reasoning to help users understand the differences between the clusters.
For example, the common features that empirical study papers share are their focus on the study method, findings, and implications.
While system papers do not have that focus: they usually concentrate on tool design, evaluation, and design recommendations instead.
Users are allowed to refine the clustering as they see fit.
For example, one might merge the ethnographic cluster with empirical studies, as they share many common features, with ethnographic study being a special case of empirical study.

\subsection{Stage 2: Abstraction}
Abstraction identifies structural commonalities among examples within each cluster.  
It synthesizes multiple examples and extracts generalizable insights, a task that can often be overwhelming for individuals due to the high cognitive load involved.

For each cluster, we instruct DeepSeek-R1 to analyze the examples and abstract the schema, which covers:
\begin{itemize}[noitemsep]
\item Components: Core elements (e.g., Method and Findings for empirical study papers).
\item Attributes: Content norms (e.g., Method should describe the empirical approach).
\item Relationships: Causal links (e.g., Findings → Implications: implications should be translated from findings).
\end{itemize}
This approach distills executable guidelines for the specific cluster, ensuring that the output not only captures the key structural dimensions (Components) but also details what makes each component effective (Attributes) and how to integrate these components into a cohesive output (Relationships).
Here is a schema that DeepSeek-R1 derives for examples within the empirical study cluster:

\begin{itemize}[noitemsep]
    \item Motivation: Introduce the research area, its relevance to HCI, and real-world significance.
        \begin{itemize}
            \item \textit{Motivation → Problem: Establishes why the research is important and what needs addressing.}
        \end{itemize}
    \item Problem: Identify what is missing in prior work or current practice.
        \begin{itemize}
            \item \textit{Problem → Method: The identified problem justifies the chosen method.}
        \end{itemize}
    \item Method: Describe the empirical approach, such as experiments or mixed methods.
        \begin{itemize}
            \item \textit{Method → Findings: The method directly leads to the results.}
        \end{itemize}
    \item Findings: Present key results, often with statistical evidence or qualitative insights.
        \begin{itemize}
            \item \textit{Findings → Implications: Results are translated into actionable insights for HCI.}
        \end{itemize}
    \item Implications: Discuss contributions to HCI research, design, policy, or practice.
\end{itemize}

Furthermore, DeepSeek-R1 is instructed to generate a table that illustrates how each example conforms to the schema. 
This enables users to assess the schema’s coherence by mapping it back to the examples. 
Users can then conveniently review the schema-example alignment and, if necessary, regenerate or modify the schema.

\begin{figure*}
\centering
\includegraphics[width=0.63\textwidth]
{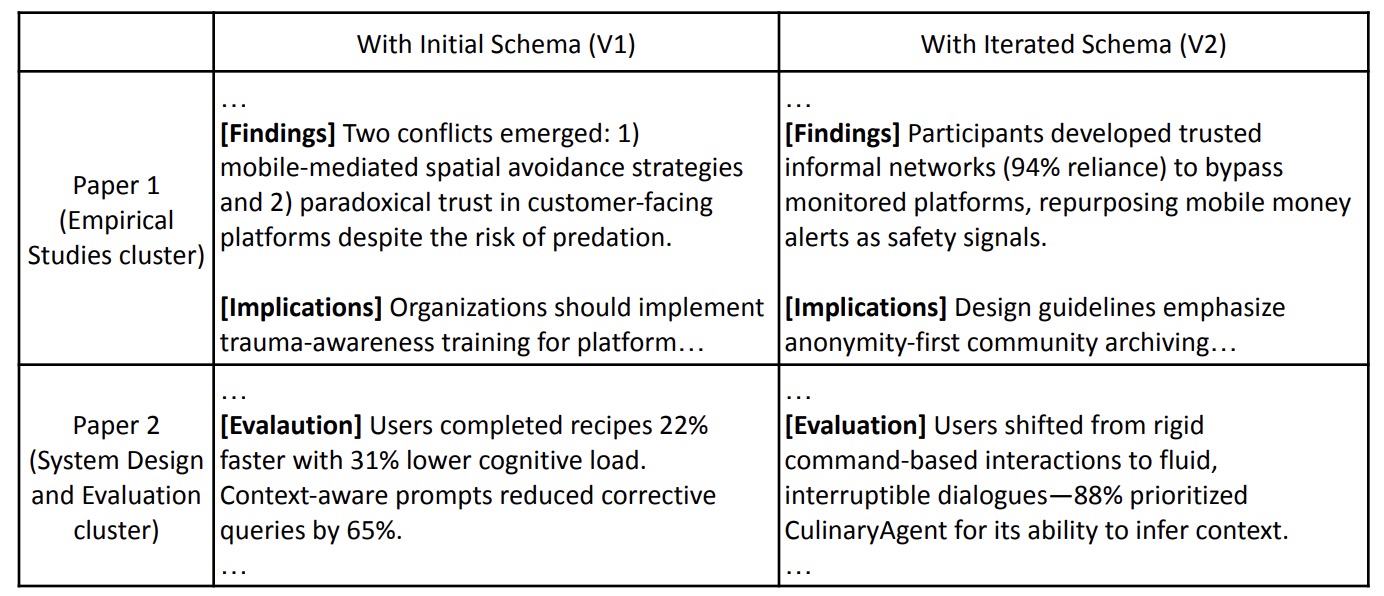}
    \caption{Case study 1 - Comparison of HCI paper abstracts generated by AI with the initial and iterated schemas.}
    \label{fig:study1}
\end{figure*}

\subsection{Stage 3: Refinement via Contrasting Examples}
The schema derived from Stage 2 is not perfect. 
For instance, while it specifies that Method should describe the empirical approach, it is not clear how to write the descriptions. 
To make the schema more precise and actionable, we employ an apply-and-test approach, iterating on the schema using contrastive learning.

Specifically, we take the initial schema from Stage 2 and prompt GPT-4 to generate HCI paper abstracts based on that schema and the paper title.
We then utilize DeepSeek-R1 to compare these generated abstracts with original human-authored ones, iterating the schema accordingly. 
For example, after one iteration, DeepSeek-R1 provided a more detailed schema for writing Method, advising users to name the empirical approach, describe the sample and duration, and connect the method to the research questions. 
As another example, the iterated schema also captures additional insights for Findings, such as emphasizing unexpected results.

\begin{itemize}[noitemsep]
    \item Method
            \begin{itemize}[noitemsep]
                \item Approach: Name the empirical approach 
                \item Sample/Duration: Include sample size, duration, and tools 
                \item Design: Link the method to the research question 
            \end{itemize}
    \item Findings
            \begin{itemize}[noitemsep]
                \item Unexpected Results: Note surprises
                \item Quantitative: Report statistical significance
                \item Qualitative: Highlight themes
            \end{itemize}
\end{itemize}

The workflow provides a compare-and-contrast view of before-and-after examples, allowing users to easily identify differences and assess the quality of the generated content. 
Users are involved in the process by evaluating whether the generated content meets their standards. 
The cycle repeats until users are satisfied.

Ultimately, users can utilize the resulting schemas as a guide to write their own HCI paper abstracts.
\section{Case Study 1: Analysis of HCI Paper Abstracts}

In this case study, we utilized Schemex to analyze the abstracts of 20 CHI'24 Best Papers to evaluate its capability for deriving actionable schemas. 
For simplicity, we accepted AI-generated answers without edits and conducted only one iteration cycle.

The workflow began with DeepSeek-R1 performing clustering, which revealed four distinct categories: Empirical Studies (6 examples), Theoretical Contributions (4), System Design \& Evaluation (6), and Ethnographic Insights (4). 
Validation against authors' pre-tagged ACM CCS concepts showed 95\% alignment (19 out of 20 papers). 
The single outlier was a paper that blended the technical design of robotics with ethnographic observations of human-animal interactions, which did not fit neatly into any cluster.

To evaluate the utility of the output schema, we conducted a blinded comparison using two randomly selected examples (one from each cluster: Empirical Studies and System Design \& Evaluation). 
GPT-4 generated paired abstracts using both the initial Stage 2 schemas and the refined Stage 3 schemas, guided by the titles of the original papers. 
An HCI expert with over five years of publication experience evaluated these abstracts through randomized pairwise comparisons and provided qualitative feedback.

The results (see Figure~\ref{fig:study1}) show marked improvements in abstracts guided by the iterated schema.
For the empirical study cluster, the iterated schema highlights that Implication should avoid ambitious claims unsupported by the study findings.
The abstract created using the iterated schema (V2) anchors the implication (``anonymity-first community archiving'') directly to findings about participants repurposing mobile money alerts to bypass monitored platforms, creating a causal link between observed behaviors and design recommendations.
In contrast, the abstract generated with the initial schema (V1) proposes ``trauma-awareness training'' as a solution but fails to connect it to the reported findings about ``spatial avoidance strategies'' and ``paradoxical trust in platforms.'' 
This creates a logical gap between findings and implications.

For the system design and evaluation cluster, the iterated schema highlights the importance of integrating qualitative context with quantitative data for effective Evaluation sentence writing.
While V1 (with initial schema) reports isolated performance statistics (22\% faster completion, 31\% reduced cognitive load, 65\% fewer corrective queries), it lacks explanatory depth about user experience drivers.
V2 (with iterated schema) presents evaluation findings by balancing quantitative metrics (88\% preference rate) with qualitative insights explaining the behavioral shift and reasons for preference.

\section{Case Study 2: Multimodal Analysis of News TikToks}

\begin{figure*}
\centering
\includegraphics[width=0.66\textwidth]
{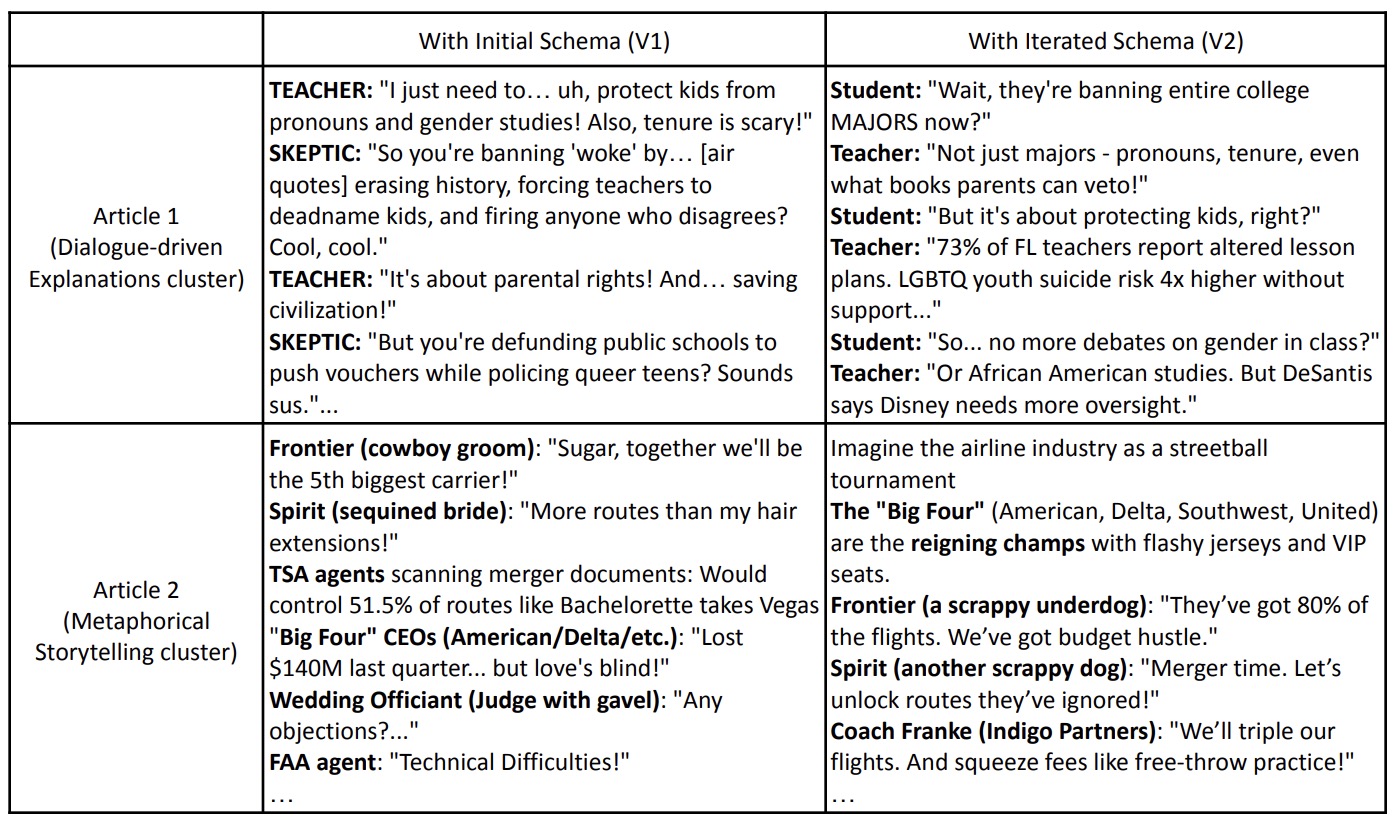}
    \caption{Case study 2 - Comparison of news TikTok scripts generated by AI with the initial and iterated schemas.}
    \label{fig:study2}
\end{figure*}

In this case study, we tested Schemex on news TikToks, a multimodal domain combining text, visual, and audio elements, which currently has less established schema knowledge. 
This builds upon our previous work, ReelFramer~\cite{reelframer}, where we manually identified schemas by analyzing examples of news TikToks.
The schema we manually derived covered three narrative framings: expository dialog, reenactment, and comedic analogy. 
It also included premises that cover characters, plot, and three key information points, among others. 
By applying Schemex to this more challenging domain, we aimed to assess whether the workflow could discover the schemas we identified through manual discovery and improve upon them.

We randomly collected 20 TikToks from The Washington Post (WAPO) for analysis, following the practice in our manual process. 
We then took preprocessing steps for multimodal analysis: for visual information, we extracted a video frame from the news TikTok every second; we then provided these frame screenshots to GPT-4V to generate captions and visual descriptions of keyframes. 
For semantic/audio information, we used Whisper to process the video and generate an audio transcript. 
After preprocessing, we obtained the visual and audio transcripts for the TikTok examples as input. 
We also included the original news articles in the input data, as they are important for understanding the creation strategies. 
Following the methodology employed in Case Study 1, we then applied Schemex to analyze these 20 WAPO news TikTok examples.

The clustering step revealed four news TikTok types: Dialogue-driven Explanations (5 examples), Metaphorical Storytelling (5), Direct Presenter \& Visual Aids (7), and Pop Culture Parody \& Memes (3). 
Validation against the researcher’s manual annotation showed 85\% alignment (17/20 TikToks). 
Two of the three outliers should have been clustered in the Pop Culture Parody \& Memes group but were incorrectly clustered elsewhere because DeepSeek-R1 did not recognize the meme. 
The other outlier does not fit into any of the clusters but is more of a blend of Pop Culture Parody and Metaphorical Storytelling. 
Compared to our findings in ReelFramer, three clusters map well, while the third cluster (Direct Presenter \& Visual Aids) provided new insights that we had not covered.

To evaluate schema utility, we conducted a blinded comparison using two randomly selected examples (one from each cluster: Dialogue-driven Explanations and Metaphorical Storytelling). 
GPT-4 generated paired news TikTok scripts using both the initial Stage 2 schemas and the refined Stage 3 schemas, guided by the original news article. 
An expert with experience in both journalism and TikTok evaluated these TikTok scripts through randomized pairwise comparisons, providing qualitative feedback.

The results (see Figure~\ref{fig:study2}) showed marked improvements in the iterated schema-guided news TikTok.
For the dialogue-driven explanation cluster, the iterated schema highlights that the characters should be relevant to the news.
While V1 relies on abstract ideological debates between generic ``Teacher'' and ``Skeptic'' personas.
The TikTok script created using the iterated schema (V2) successfully anchors its critique of Florida’s education policies through role-specific characters (student/teacher) that mirror real-world stakeholders affected by the legislation, making the exchanges more natural. 
For the metaphorical storytelling cluster, the iterated schema highlights that the script should establish a single cohesive metaphor rather than mixing multiple disjoined references.
While V1 (with initial schema) incorporates disparate elements that include wedding characters (groom, bride, wedding officiant), TSA, and FAA characters, V2 (with iterated schema) adheres to a unified sports analogy, mapping news characters to relevant, identifiable roles such as VIP players, underdog teams, and coach.
This approach makes the narrative more comprehensible and cohesive.

\section{Concluding Thoughts}

Schemex is designed to help users extract actionable schemas from examples. 
Through our previous experiences with manually deriving schemas, we have identified challenges in clustering examples and evaluating schemas, especially as the number and complexity of examples increase. 
During the initial evaluation of Schemex, we found that AI, particularly recent reasoning models, is capable of performing clustering, significantly reducing the cognitive load on users. 
Furthermore, to streamline the schema evaluation process, Schemex incorporates an apply-iterate cycle. 
Current AI reasoning models can conduct in-depth analyses by comparing poor and strong examples, which supports meaningful iterations and alleviates the burden on users.

Overall, we view Schemex as a tool that augments human cognition through clustering, abstraction, and contrastive learning. 
AI serves as a starting point to relieve humans of the initial heavy lifting, while humans are involved in evaluation. 
Looking ahead, we plan to refine how we construct the workflow by integrating key components like abstraction and iteration in a more flexible manner. 
This includes making more intelligent decisions about low-level operations, such as determining the necessary number of iterations, allowing users to concentrate on high-level thinking.

\bibliographystyle{ACM-Reference-Format}
\bibliography{sample-base}

\end{document}